\documentclass[conference, compsoc]{IEEEtran}
\IEEEoverridecommandlockouts
\usepackage{cite}
\usepackage{amsmath}
\usepackage{amsfonts}
\usepackage{multirow}

\usepackage{graphicx}
\usepackage{textcomp}
\usepackage[caption=false]{subfig}

\usepackage{breqn}
\usepackage{booktabs,ragged2e}
\newcommand*\xor{\oplus}

\usepackage[flushleft]{threeparttable}

\newcommand\ncoverline[1]{\mkern1mu\overline{\mkern-1mu#1\mkern-1mu}\mkern1mu}

\begin{document}
	\title{On Approximate 8-bit Floating-Point Operations Using Integer Operations
	\thanks{This work was financed by the ELLIIT strategic research environment through the project D3 ACRE -- Approximate Computing Reducing Energy.}
}


\author{\IEEEauthorblockN{1\textsuperscript{st} Theodor Lindberg}
	\IEEEauthorblockA{Dept. of Electrical Engineering \\
		Linköping University\\
		SE-581 83 Linköping, Sweden \\
		\texttt{theodor.lindberg@liu.se}}
	\and
	\IEEEauthorblockN{2\textsuperscript{nd} Oscar Gustafsson}
	\IEEEauthorblockA{Dept. of Electrical Engineering\\
		Linköping University\\
		SE-581 83 Linköping, Sweden \\
		\texttt{oscar.gustafsson@liu.se}}
}
	
	\maketitle

	\begin{abstract}
		In this work, approximate eight-bit floating-point operations performed using simple integer operations is discussed.
		For two-bit mantissa formats, faithful rounding can always be obtained for the considered operations. For all operations, correctly rounded results can be obtained for different rounding modes, either directly or by adding a conditional carry in.
		For three-bit mantissa formats, faithful rounding can be sometimes be obtained directly, while for other operations a conditional carry in must be added.
		Correctly rounded results can be obtained for most operations and rounding modes using slightly more complicated expressions for the carry in.
		Hardware implementation results for multiplication using both standard cell and FPGA technology are presented illustrating the potential benefit of integer computation.
		Especially for FPGA, significant resource savings are obtained.
	\end{abstract}

	
	
	\section{Introduction} \label{sec:introduction}
	Eight-bit floating-point (FP8) formats is an emerging topic in error resilient applications \cite{Park2022SEB,noune20228bit,micikevicius2022fp8}. It has recently even been adapted by some of the leading edge architectures specializing in machine learning -- such as the H100 Tensor Core from NVIDIA\cite{NVIDIA2023H100} and the Gaudi 2 AI accelerator from Intel\cite{Intel2023Gaudi2}.
	Other short floating-point formats, such as FP16, has also seen fit in Internet of Things (IoT) devices, where low energy consumption is key\cite{Mach2018Transprecision, Aguilera2018Half}. 
	Together with the increase of IoT devices, an incentive for edge computing has also emerged.
	Edge computing solutions are needed in applications with low latency, such as 6G, to cope with the strict requirements.
	Ideas about deploying CNN networks on IoT devices are also becoming reality, and something that will play an eminent role in 6G networks\cite{Zawish2023Energy}.
	
	Floating-point operations are however still expensive.
	While using FP8 has many advantages in lower precision applications, such as shorter latency, smaller memory footprint, and reduced energy consumption, very few platforms support FP8 operations as of today.
	Let alone, it is not rare for smaller microcontrollers to not include a floating-point unit (FPU) at all. 
	Some have for example implemented pseudo FP16 operations on eight-bit microcontrollers\cite{Roshchupkin2011Neural}.
	
	Additionally, research points towards multiple FP8 formats are needed within the same application\cite{Park2022SEB, noune20228bit}.
	The two most prominent FP8 formats being E5M2 and E4M3, where the numbers indicates how many exponent and mantissa bits they use respectively. Both being supported on the H100 Tensor Core and Gaudi 2\cite{NVIDIA2023H100, Intel2023Gaudi2}.
	
	In this work, the implementation of approximate FP8 operations using integer arithmetic operations is considered. This allows implementation of approximate FP8-operations using integer vector/SIMD instruction, something that can be efficient when there is no hardware support for FP8. It is shown that for certain rounding modes, correctly rounded results can be obtained. Also, it is shown that by forming a carry-in term, most operations and rounding-modes can be supported. Expressions for these carry-in terms are provided. Implementation results for FP8 multiplication show that the integer operation approach is attractive, especially for FPGA implementation.
	
	
	\section{Approximate Floating-Point Operations in the Logarithmic Domain}
	A normal floating-point number $x$ can be expressed using the triplet $(s_x, e_x, m_x)$ such that
	\begin{equation}\label{eq:fp-definition}
		x = (-1)^{s_x} (1 + m_x) 2^{e_x},
	\end{equation}
	where $s_x \in \{0, 1\}$, $0 \leq m_x < 1$, and $e_{\min} \leq e_x \leq e_{\max}$.
	
\begin{figure} 
		\centering
		\subfloat[The E5M2 format.\label{fig:e5m2-format}]{%
			\includegraphics{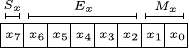}}
		\hfill
		\subfloat[The E4M3 format.\label{fig:e4m3-format}]{%
			\includegraphics{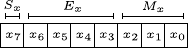}}
		\caption{Binary representation of the considered FP8 formats.}
		\label{fig:fp8-formats}
	\end{figure}

	The binary interchange formats usually follow the IEEE 754-2019 Standard\cite{754Std2019} where the aforementioned triplet is stored as three fields; a sign bit $S_x$, follow by a $W_E$-bit biased exponent ${E_x=e_x + b}$, and a $(p-1)$-bit trailing integral significand $M_x$, where the bias $b$ is equal to $2^{W_E-1} - 1$ and $M_x$ equals $m_x \cdot 2^{p-1}$\cite{Muller2009Handbook}.
	Figures~\ref{fig:e5m2-format}~and~\ref{fig:e4m3-format} illustrates the E5M2 and E4M3 binary interchange formats as defined in \cite{micikevicius2022fp8}.
	
	A floating-point number of such form can be converted to a sign-magnitude fixed-point number $\hat{X}$ in the LNS domain using Mitchell's approximation\cite{Mitchell1962Computer}, as explained in \cite{gustafsson2021approximate}, by
	\begin{equation}\label{eq:to-log}
		(-1)^{s_x}\log_2|x| \approx X - B = \hat{X},
	\end{equation}
	where $X$ is the binary representation of $x$ and $B$ equals $b \ll (p-1)$. The operation is thus very cheap and the conversion back from LNS is simply interpreting
	\begin{equation}\label{eq:from-log}
		\hat{X} + B
	\end{equation}
	as a floating-point number.
	The fixed-point number $\hat{X}$ consists of a sign bit $S_x$, an integer part $E_x$, and a fractional part $M_x$. Similarly, the constant $B$ can be interpreted as the fixed-point representation of the bias $b$.
	
	Table~\ref{tab:lns-expressions} shows a collection of approximate operations presented in \cite{gustafsson2021approximate}, which serve as a starting point in this work.
	The constant $B$ depends on the format and equals ${(15\ll 2)=60=\mathtt{0x3c}}$ for E5M2, and ${(7\ll 3)=56=\mathtt{0x38}}$ for E4M3. Negative constants are written as eight-bit signed numbers using two's complement.
	\begin{table}
		\caption{Approximate Operations in LNS}
		\label{tab:lns-expressions}
		\centering
		\begin{tabular}{ c | c c }
			Operation & Mathematical notation & Expression in LNS \\ 
			\hline
			Multiplication   & $x \times y$ & $X + Y - B$  \\
			Square   & $x^2$ & $X \ll 1 - B$  \\
			Division   & $x / y$ & $X - Y + B$  \\
			Reciprocal   & $1/x$ & $-X + 2B$  \\
			Square-root   & $\sqrt{x}$ & $X \gg 1 + B/2$  \\
			Reciprocal square-root   & $1/\sqrt{x}$ & $-X \gg 1 + 3B/2$  \\
		\end{tabular}
	\end{table}

	\section{Error Analysis and Compensations}\label{sec:deciding-error-comp}
	As the operations are performed in the LNS domain and the aim is to achieve different rounding modes, it makes sense to reason in terms of number of units in the last place (ulp). While there exists several definitions of the ulp function\cite{muller2006ulp}, this work uses a slight modification of the definition from Goldberg \cite{Goldberg1991What}. The definition from Goldberg states if the floating-point $d_0.d_1\dots d_{p-1}\beta^e$ is used to represent a number $z$, then its error is by
	\begin{equation}\label{eq:goldberg-ulp}
		|d_0.d_1\dots d_{p-1}\ - z/\beta^e|\beta^{p-1}
	\end{equation}
	ulps, where $\beta$ is the radix.
	However, to differentiate over and under approximations the absolute error is not used.
	While $z$ may be any real number, to achieve a particular rounding mode the reference is the quantized value of $z$ after using the rounding mode of interest.
	The result from an operation before quantization is referred to as the mathematically exact result.
	
	
	Seven different rounding modes are taken into consideration in this work:
	\begin{enumerate}
		\item RN$_e$: round to closest, ties to even,
		\item RN$_a$: round to closest, ties to away,
		\item RN$_z$: round to closest, ties to zero,
		\item RU: round towards positive infinity,
		\item RD: round towards negative infinity,
		\item RZ: round towards zero,
		\item faithful rounding.
	\end{enumerate}
	The six first rounding modes are defined by the IEEE 754-2019 Standard\cite{754Std2019}, where RN$_e$ is the default mode.
	RN$_z$ was introduced in the revision of the 2008 Standard, and was added specifically for augmented operations\cite{754Std2019, Riedy2018Augmented}.
	
	A result is considered faithful if $\operatorname{RD}(x)$ or $\operatorname{RU}(x)$ is always returned\cite{Muller2009Handbook}, faithful rounding is therefore defined in this work as a mode, $\circ$,  that fulfills
	\begin{equation}
		\operatorname{RD}(x) \leq \circ (x) \leq \operatorname{RU}(x).
	\end{equation} 

	Due to the similarities between the modes, many expressions for the error compensations are likely to be shared.
	Remember, the only difference between the modes with correct rounding is the handling of tie breaks, and the difference between the modes with direct rounding is the handling of sign. 
	\subsection{E5M2}
	This section studies the considered operations for the E5M2 format.
	
	\subsubsection{Multiplication}
\begin{figure}[]
	\centering
	\includegraphics[]{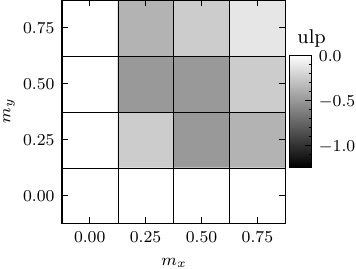}
	\caption{Error in ulp for approximate E5M2 multiplication using (\ref{eq:e5m2-mul}) compared to the mathematically exact result.}
	\label{fig:e5m2-mul-ulp-math}
\end{figure}

	The multiplication is approximated as
	\begin{equation}
		X + Y - B = X + Y + \mathtt{0xc4}. \label{eq:e5m2-mul}
	\end{equation}
	Figure~\ref{fig:e5m2-mul-ulp-math} shows the error compared to the mathematically exact result.
	As the error is always positive and at most $1/2$~ulp, both RN$_z$ and RZ are obtained without any error correction.

\begin{figure}[]
		\centering
		\includegraphics[]{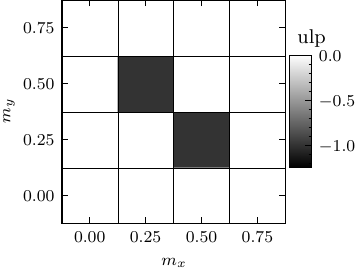}
		\caption{Error in ulp for approximate E5M2 multiplication using (\ref{eq:e5m2-mul}) compared to RN$_e$.}
		\label{fig:e5m2-mul-ulp-rne}
	\end{figure}

	To study the other rounding modes further, the quantized error for each mode is calculated. Figure~\ref{fig:e5m2-mul-ulp-rne} shows the error with regards to RN$_e$. From here it can be seen that one ulp must be added when $m_x$ equals $0.25$ and $m_y$ equals $0.5$, or vice versa.
	The ulp can be added via a conditional carry in, $c_{\mathrm{in}}$, as
	\begin{equation}\label{eq:cin-e5m2-mul-rne}
		c_{\mathrm{in}}=x_{0}  y_{1}  \ncoverline x_{1}  \ncoverline y_{0} + x_{1}  y_{0}  \ncoverline x_{0}  \ncoverline y_{1}.
	\end{equation}
	
	The carry-in expression for RN$_a$ only differs from (\ref{eq:cin-e5m2-mul-rne}) at the other tie break, i.e., when both $m_x$ and $m_y$ equal $0.5$, leading to
	\begin{equation}\label{eq:cin-e5m2-mul-rna}
		c_{\mathrm{in}}=x_{0}  y_{1}  \ncoverline x_{1}  \ncoverline y_{0} + x_{1}  y_{0}  \ncoverline x_{0}  \ncoverline y_{1} + x_{1}  y_{1}  \ncoverline x_{0}  \ncoverline y_{0}.
	\end{equation}
	
	The directed modes RU and RD are correct when one of the operands are zero. For the remaining cases, a carry in should be added depending on the sign or the result, $S_r = S_x \xor S_y$, as 
	\begin{equation}\label{eq:cin-e5m2-mul-ru}
		c_{\mathrm{in}}=\ncoverline S_r  \left(x_{0} + x_{1}\right)  \left(y_{0} + y_{1}\right)
	\end{equation}
	for RU and 
	\begin{equation}\label{eq:cin-e5m2-mul-rd}
		c_{\mathrm{in}}=S_r  \left(x_{0} + x_{1}\right)  \left(y_{0} + y_{1}\right)
	\end{equation}
	for RD.

	\subsubsection{Square}
	The square is approximated as
	\begin{equation}
		X \ll 1 - B = X \ll 1 + \mathtt{0xc4},
	\end{equation}
	and can be deducted from multiplication with $m_y = m_x$. Hence, the rounding modes RN$_e$, RN$_z$, RD, and RZ are all achieved without any carry in, i.e. $c_{\mathrm{in}}=0$.
	
	The edge case for RN$_a$ is when $m_x$ equals $0.5$, which gives the boolean expression 
	\begin{equation}
		c_{\mathrm{in}}=x_{1}  \ncoverline x_{0} \label{eq:cin-e5m2-sq-rna}
	\end{equation}
	for the carry in.
	
	RU is only correct when $m_x=0$, so the carry in therefore becomes
	\begin{equation}
		c_{\mathrm{in}}=x_{0} + x_{1}. \label{eq:cin-e5m2-sq-ru}
	\end{equation}

	\subsubsection{Division}
\begin{figure}[]
		\centering
		\includegraphics[]{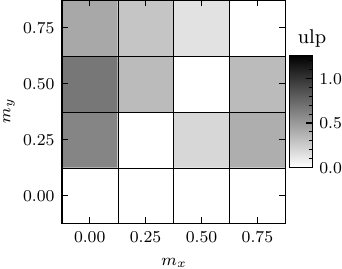}
		\caption{Error in ulp for approximate E5M2 division using (\ref{eq:e5m2-div}), compared to the mathematically exact result.}
		
		\label{fig:e5m2-div-ulp-error-math-orig}
	\end{figure}

\begin{figure}[]
		\centering
		\includegraphics[]{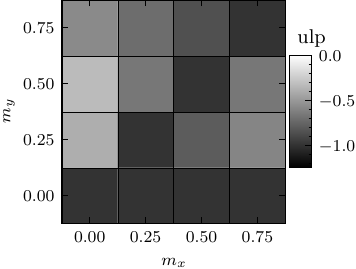}
		\caption{Error in ulp for approximate E5M2 division using (\ref{eq:e5m2-div-dec}), compared to the mathematically exact result.}
		\label{fig:e5m2-div-ulp-error-math}
	\end{figure}

	Unlike for the previous cases, the division approximation
	\begin{equation}
		X - Y + B = X - Y + \mathtt{0x3c}, \label{eq:e5m2-div}
	\end{equation}
	gives an over-approximation, as shown in Fig.~\ref{fig:e5m2-div-ulp-error-math-orig}.
	However, as the error is in $[0, 1\,\mathrm{ulp}]$ it is possible to decrease the constant such that an under-approximation is obtained, see Fig.~\ref{fig:e5m2-div-ulp-error-math}, and a conditional carry in can compensate the error.
	

	The new expression with a conditional carry in thus becomes
\begin{equation}
	X - Y + B - 1 + c_{\mathrm{in}} = X - Y + \mathtt{0x3b}  + c_{\mathrm{in}}. \label{eq:e5m2-div-dec}
\end{equation}

	The modes with correct rounding, RN$_e$, RN$_a$, and RN$_z$, can all share the same carry in expression
	\begin{equation}\label{eq:cin-e5m2-div-rne-rna-rnz}
		c_{\mathrm{in}}=x_{0} + x_{1} + y_{0}  y_{1} + \ncoverline y_{0}  \ncoverline y_{1},
	\end{equation}
	while the other modes have some variations.
	RZ, RU, and RD all share the term
		\begin{equation}\label{eq:cin-e5m2-div-rz}
		c_{\mathrm{in}}=\ncoverline y_{0}  \ncoverline y_{1} + x_{0}  \ncoverline x_{1}  \ncoverline y_{1} + x_{1}  \ncoverline x_{0}  \ncoverline y_{0} + x_{0}  x_{1}  y_{0}  y_{1},
	\end{equation}
	 but RU and RD also consider the resulting sign $S_r = S_x \xor S_y$ as
	\begin{equation}\label{eq:cin-e5m2-div-ru}
		c_{\mathrm{in}}=\ncoverline S_r + \ncoverline y_{0}  \ncoverline y_{1} + x_{0}  \ncoverline x_{1}  \ncoverline y_{1} + x_{1}  \ncoverline x_{0}  \ncoverline y_{0} + x_{0}  x_{1}  y_{0}  y_{1}
	\end{equation}
	and
	\begin{equation}\label{eq:cin-e5m2-div-rd}
		c_{\mathrm{in}}=S_r + \ncoverline y_{0}  \ncoverline y_{1} + x_{0}  \ncoverline x_{1}  \ncoverline y_{1} + x_{1}  \ncoverline x_{0}  \ncoverline y_{0} + x_{0}  x_{1}  y_{0}  y_{1},
	\end{equation}
	respectively.

	\subsubsection{Reciprocal}
	Reciprocal is the same as division with numerator equal to 1, so the expression 
	\begin{equation}
		-X + 2B = -X + \mathtt{0x88}
	\end{equation}
	results in an over approximation, and the constant must therefore be decreased to $\mathtt{0x87}$ leading to
	\begin{equation}
		-X + 2B - 1 + c_{\mathrm{in}}= -X + \mathtt{0x87} + c_{\mathrm{in}} 
	\end{equation}
	
	RN$_e$, RN$_a$, RN$_z$ must be compensated for $m_x=0$ and $m_x=0.75$, so the carry in expression thus becomes
	\begin{equation}\label{eq:cin-e5m2-recip-rne-rna-rnz}
		c_{\mathrm{in}}=x_{0}  x_{1} + \ncoverline x_{0}  \ncoverline x_{1}.
	\end{equation}
	For RZ, the edge case is only when $m_x=0$, which gives the carry in expression
	\begin{equation}\label{eq:cin-e5m2-recip-rz}
		c_{\mathrm{in}}=\ncoverline x_{0}  \ncoverline x_{1}.
	\end{equation}
	RU and RD again only differ at the sign, $S_r = 0 \xor S_x = x_7$, as shown in (\ref{eq:cin-e5m2-recip-ru}) and (\ref{eq:cin-e5m2-recip-rd}) respectively.
	\begin{equation}\label{eq:cin-e5m2-recip-ru}
		c_{\mathrm{in}}=x_{7} + \ncoverline x_{0}  \ncoverline x_{1},
	\end{equation}
	\begin{equation}\label{eq:cin-e5m2-recip-rd}
		c_{\mathrm{in}}=\ncoverline x_{7} + \ncoverline x_{0}  \ncoverline x_{1}
	\end{equation}

	\subsubsection{Square-Root}\label{sec:e5m2-sqrt}
	The square-root approximate is determined as
	\begin{equation}\label{eq:e5m2-sqrt}
		X \gg 1 + \frac{B}{2} = X \gg 1 + \mathtt{0x1e}.
	\end{equation}
	This gives correct rounding for RN$_e$, RN$_a$, and RN$_z$.
	However, (\ref{eq:e5m2-sqrt}) will under-approximate RU, and over-approximate the other directed modes.
	It is therefore not possible to select a constant such that all rounding modes can be achieved by solely selecting the carry in. RU can be compensated by forwarding the last bit to the carry in,
	\begin{equation} \label{eq:cin-e5m2-sqrt-ru}
		c_{\mathrm{in}}=x_{0}.
	\end{equation}
	The error of (\ref{eq:e5m2-sqrt}) repeats itself every other exponent for other formats, as the last bit of the exponent field will shift down to the mantissa.
	However since there are so few mantissa bits, this property does not appear here.
	
	\subsubsection{Reciprocal Square-Root} 
	The approximate reciprocal square-root,
	\begin{equation}
		-X \ll 1 + \frac{3B}{2} = -X \ll 1 +  \mathtt{0x5a},
	\end{equation}
	yields a slight over-approximation, however due to quantization the error compensation is identical to that of the square-root (\ref{eq:e5m2-sqrt}). 
	
	\subsection{E4M3}
	The method for deciding error compensation for E4M3 is similar to that of E5M2, and is therefore kept more brief.
	
	It is here important to note that even though the boolean expressions for the carry ins for E4M3 will be large, they will in fact map very well to FPGA architectures. This is because the expression only depends on six variables at most, i.e. the two mantissas, and can therefore fit inside a single LUT on many architectures. Furthermore, the synthesis tool may even combine the first full adder and the carry in as they both solely depend on $m_x$ and $m_y$.
	
	\subsubsection{Multiplication}\label{sec:e4m3-mul}
\begin{figure}
		\centering
		\includegraphics[]{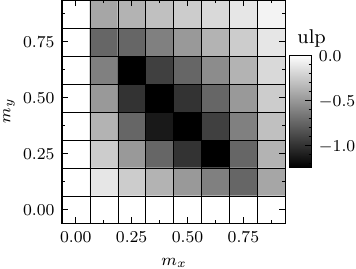}
		\caption{Error in ulp for the approximate E4M3 multiplication in (\ref{eq:e4m3-mul}) compared to the mathematically exact result.}
		\label{fig:e4m3-mul-ulp-math}
	\end{figure}

	For multiplication, the approximation
	\begin{equation}
		X + Y - B = X + Y + \mathtt{0xc8}, \label{eq:e4m3-mul}
	\end{equation}
	is used. 
	Figure~\ref{fig:e4m3-mul-ulp-math} shows the error compared to the mathematically correct result, and as the error is at most $-1.5$ ulp, none of the rounding modes are obtained directly.
	Consequently, RU and RD are not achievable. 
	The carry in expression for RN$_e$ however becomes
	\begin{dmath}\label{eq:cin-e4m3-mul-rne}
		c_{\mathrm{in}}=x_{0}  y_{2}  \ncoverline x_{2}  \ncoverline y_{0} + x_{0}  y_{2}  \ncoverline x_{2}  \ncoverline y_{1} + x_{1}  y_{2}  \ncoverline x_{2}  \ncoverline y_{0} + x_{1}  y_{2}  \ncoverline x_{2}  \ncoverline y_{1} + x_{2}  y_{0}  \ncoverline x_{0}  \ncoverline y_{2} + x_{2}  y_{0}  \ncoverline x_{1}  \ncoverline y_{2} + x_{2}  y_{1}  \ncoverline x_{0}  \ncoverline y_{2} + x_{2}  y_{1}  \ncoverline x_{1}  \ncoverline y_{2} + x_{2}  y_{2}  \ncoverline x_{1}  \ncoverline y_{1} + x_{0}  x_{1}  y_{1}  \ncoverline x_{2}  \ncoverline y_{2} + x_{1}  y_{0}  y_{1}  \ncoverline x_{2}  \ncoverline y_{2},
	\end{dmath}
	while RN$_a$ must correct a couple of additional tie break:
	\begin{dmath}\label{eq:cin-e4m3-mul-rna}
		c_{\mathrm{in}}=x_{0}  y_{2}  \ncoverline x_{1}  \ncoverline y_{1} + x_{0}  y_{2}  \ncoverline x_{2}  \ncoverline y_{0} + x_{1}  y_{1}  \ncoverline x_{0}  \ncoverline y_{2} + x_{1}  y_{1}  \ncoverline x_{2}  \ncoverline y_{0} + x_{1}  y_{1}  \ncoverline x_{2}  \ncoverline y_{2} + x_{1}  y_{2}  \ncoverline x_{2}  \ncoverline y_{1} + x_{2}  y_{0}  \ncoverline x_{0}  \ncoverline y_{2} + x_{2}  y_{0}  \ncoverline x_{1}  \ncoverline y_{1} + x_{2}  y_{1}  \ncoverline x_{1}  \ncoverline y_{2} + x_{2}  y_{2}  \ncoverline x_{0}  \ncoverline x_{1}  \ncoverline y_{0} + x_{2}  y_{2}  \ncoverline x_{0}  \ncoverline y_{0}  \ncoverline y_{1}.
	\end{dmath}
	As the approximation gives a slightly smaller result, the tie breaks for RN$_z$ will be correct and the carry in expression becomes
	\begin{dmath}\label{eq:cin-e4m3-mul-rnz}
		c_{\mathrm{in}}=x_{1}  y_{2}  \ncoverline x_{2}  \ncoverline y_{0} + x_{1}  y_{2}  \ncoverline x_{2}  \ncoverline y_{1} + x_{2}  y_{1}  \ncoverline x_{0}  \ncoverline y_{2} + x_{2}  y_{1}  \ncoverline x_{1}  \ncoverline y_{2} + x_{2}  y_{2}  \ncoverline x_{1}  \ncoverline y_{1} + x_{0}  x_{1}  y_{1}  \ncoverline x_{2}  \ncoverline y_{2} + x_{0}  x_{2}  y_{0}  \ncoverline x_{1}  \ncoverline y_{2} + x_{0}  y_{0}  y_{2}  \ncoverline x_{2}  \ncoverline y_{1} + x_{0}  y_{1}  y_{2}  \ncoverline x_{2}  \ncoverline y_{0} + x_{1}  x_{2}  y_{0}  \ncoverline x_{0}  \ncoverline y_{2} + x_{1}  y_{0}  y_{1}  \ncoverline x_{2}  \ncoverline y_{2}.
	\end{dmath} 
	Lastly, the boolean expression for obtaining RZ is
	\begin{dmath}\label{eq:cin-e4m3-mul-rz}
		c_{\mathrm{in}}=x_{1}  y_{2}  \ncoverline x_{0}  \ncoverline x_{2}  \ncoverline y_{1} + x_{1}  y_{2}  \ncoverline x_{2}  \ncoverline y_{0}  \ncoverline y_{1} + x_{2}  y_{1}  \ncoverline x_{0}  \ncoverline x_{1}  \ncoverline y_{2} + x_{2}  y_{1}  \ncoverline x_{1}  \ncoverline y_{0}  \ncoverline y_{2} + x_{0}  x_{1}  y_{0}  y_{1}  \ncoverline x_{2}  \ncoverline y_{2} + x_{2}  y_{2}  \ncoverline x_{0}  \ncoverline x_{1}  \ncoverline y_{0}  \ncoverline y_{1}.
	\end{dmath}
	
	Faithful rounding can be obtained in several ways, one is to check if $m_x$ and $m_y$ are greater than $0$ which gives the carry expression
	\begin{equation}\label{eq:cin-e4m3-mul-faith}
		c_{\mathrm{in}}=(x_2 + x_1 + x_0)(y_2 + y_1 + y_0).
	\end{equation}
	\subsubsection{Square}
	From using the approximation
	\begin{equation}
		X \ll 1 - B = X \ll 1 + \mathtt{0xc8},
	\end{equation}
	the carry in compensation for RN$_e$ and RN$_z$ becomes
	\begin{dmath}\label{eq:cin-e4m3-sq-rne-rnz}
		c_{\mathrm{in}}=x_{2}  \ncoverline x_{1} + x_{0}  x_{1}  \ncoverline x_{2},
	\end{dmath}
	while RN$_a$ has a slightly smaller expression of
	\begin{dmath}\label{eq:cin-e4m3-sq-rna}
		c_{\mathrm{in}}=x_{1}  \ncoverline x_{2} + x_{2}  \ncoverline x_{1}
	\end{dmath}
	as more cases must be compensated for.
	RU will not work since the error will be $-1.5$~ulp at most (as noted in Section~\ref{sec:e4m3-mul}), but the same expression for RN and RZ can be share:
	\begin{dmath}\label{eq:cin-e4m3-sq-rn-rz}
		c_{\mathrm{in}}=x_{0}  x_{1}  \ncoverline x_{2} + x_{2}  \ncoverline x_{0}  \ncoverline x_{1}.
	\end{dmath}

	Faithful rounding can be achieved by adding an extra ulp when $m_x$ equals $0.375$ or $0.5$ by
	\begin{equation}\label{eq:cin-e4m3-sq-faith}
		c_{\mathrm{in}}=x_2 \ncoverline x_1 \ncoverline x_0 + \ncoverline x_2 x_1 x_0.
	\end{equation}

	\subsubsection{Division}
	Similar to the case for E5M2, the constant must be decremented such that the carry in can compensate the error. The division for E4M3 will thus be performed as
	\begin{equation}
		X - Y + B - 1 + c_{\mathrm{in}} = X - Y + \mathtt{0x37} + c_{\mathrm{in}}.
	\end{equation}
	
	The three correctly rounded modes, RN$_e$, RN$_a$, and RN$_z$, share the same expression 
	\begin{dmath}\label{eq:cin-e4m3-div-rne-rna-rnz}
		c_{\mathrm{in}}=x_{0}  x_{1}  \ncoverline x_{2} + x_{1}  \ncoverline x_{2}  \ncoverline y_{2} + x_{2}  y_{1}  y_{2} + x_{2}  \ncoverline x_{0}  \ncoverline x_{1} + x_{2}  \ncoverline x_{1}  \ncoverline y_{1} + y_{0}  y_{1}  y_{2} + \ncoverline y_{0}  \ncoverline y_{1}  \ncoverline y_{2} + x_{0}  \ncoverline x_{1}  \ncoverline y_{1}  \ncoverline y_{2} + x_{2}  y_{0}  y_{2}  \ncoverline x_{0}
	\end{dmath}
	for the carry in.
	No error compensation can be found for the directed modes however, as the error is $\pm1$ ulp.
	
	The error plot for division in E4M3 looks similar to that of E5M2, Fig.~\ref{fig:e5m2-div-ulp-error-math}, and from there it is possible to see that faithful rounding can be achieved by adding the carry in when $m_y=0$ and when $m_x=m_y$. This can be realized by the boolean expression
	\begin{equation}\label{eq:cin-e4m3-div-faith}
		c_{\mathrm{in}}=\ncoverline y_2 \ncoverline y_1 \ncoverline y_0 + \overline{(x_2\xor y_2)}\,\overline{(x_1\xor y_1)}\, \overline{(x_0\xor y_0)}.
	\end{equation}
	\subsubsection{Reciprocal}
	Also here must the constant be decremented as
	\begin{equation}
		-X + 2B -1 + c_{\mathrm{in}} = -X + \mathtt{0x6f} + c_{\mathrm{in}}.
	\end{equation}
	
	The directed rounding modes cannot be obtained here either, since the error has two different signs, just as before.
	The carry in expression for the correctly rounded modes on the other hand is
	\begin{dmath}\label{eq:cin-e4m3-recip-rne-rna-rnz}
		c_{\mathrm{in}}=x_{0}  x_{1}  x_{2} + \ncoverline x_{0}  \ncoverline x_{1}  \ncoverline x_{2}.
	\end{dmath}
	
	As the reciprocal is just a sub case of division, faithful rounding can be obtained by simply adding a carry in when $m_x=0$ by
	\begin{equation}\label{eq:cin-e4m3-recip-faith}
		c_{\mathrm{in}} = \ncoverline x_2 \ncoverline x_1 \ncoverline x_0.
	\end{equation}
	\subsubsection{Square-Root}
	The approximation
	\begin{equation}\label{eq:e4m3-sqrt}
		X \gg 1 + \frac{B}{2}-1 + c_{\mathrm{in}} = X \gg 1 + \mathtt{0x1b} + c_{\mathrm{in}}.
	\end{equation}
	is used. It will however under-approximate RN$_e$, RN$_a$, and RN$_z$ when the least significant bit of the exponent is $1$, the expression thus becomes
	\begin{equation}\label{eq:cin-e4m3-sqrt-rne-rna-rnz}
		c_{\mathrm{in}}=\ncoverline x_3 + x_{0} + x_{1} + x_{2}.
	\end{equation}	
	The fact that the error depends on $x_3$ is due to the right shift mentioned in Section~\ref{sec:e5m2-sqrt}.
	
	A maximum error of $-2$ ulps is obtained when comparing with RU, so compensating with a carry in is thus not possible. However the rounding modes RD and RZ are possible to obtain by using
	\begin{dmath}\label{eq:cin-e4m3-sqrt-rn-rz}
		c_{\mathrm{in}}= x_{3} x_{0} + \ncoverline x_{3}\left( x_{0}  \ncoverline x_{1} + x_{0}  \ncoverline x_{2} + \ncoverline x_{1}  \ncoverline x_{2}\right)
	\end{dmath}
	as the expression for the carry in.
	RD and RZ are naturally the same as the square-root will by definition produce a positive number.
	
	\subsubsection{Reciprocal Square-Root}
	Performing the reciprocal square-root approximation as
	\begin{equation}
		-X \ll 1 + \frac{3B}{2} = -X \ll + \mathtt{0x54},
	\end{equation}
	will over-approximate all rounding modes except RU. Hence, the expression used is
		\begin{equation}
		-X \ll 1 + \frac{3B}{2}  - 1 + c_{\mathrm{in}} = -X \ll 1 + \mathtt{0x53} + c_{\mathrm{in}}.
	\end{equation}
	
	The carry in expression for RN$_e$, RN$_a$, and RN$_z$ becomes
	\begin{dmath}\label{eq:cin-e4m3-recip-sqrt-rne-rna-rnz}
		c_{\mathrm{in}}=x_3\ncoverline x_{1}  \ncoverline x_{2} + \ncoverline x_3 x_{1}  x_{2} + x_0, 
	\end{dmath}
	and, again, notice that the error depends on the least significant bit of the exponent, $x_3$.
	
	For RN and RZ the expression becomes
	\begin{equation}\label{eq:cin-e4m3-recip-sqrt-rn-rz}
		c_{\mathrm{in}}=x_3\ncoverline x_{1}  \ncoverline x_{2} + \ncoverline x_3 x_{0}  x_{1}  x_{2}.
	\end{equation}

	\subsection{Summary of Obtained Expressions}
	Tables~\ref{tab:e5m2-expressions} and~\ref{tab:e4m3-expressions} summarizes the obtained expressions for E5M2 and E4M3 respectively. Where the carry in expressions are too long to fit inside the table the equation is instead referenced. For rounding modes where correcting the error with just the carry in was not possible a dash is written. Note that in some cases all possible rounding modes was possible to obtain, but the same constant can not be used.
	\begin{table*}
		\centering
		\begin{threeparttable}
		\caption{Summary of Suggested Expressions for E5M2.}
		\label{tab:e5m2-expressions}
		\begin{tabular}{c|c|*{7}{c}}
			 & & \multicolumn{7}{c}{Boolean expressions for  carry in, $c_{\mathit{in}}$\tnote{a}} \\
			Operation   & Integer arithmetic expression & RN$_e$ & RN$_a$ & RN$_z$ & RU & RD & RZ & Faithful \\
			\hline
			$x\times y$ & $X + Y + \mathtt{0xc4} + c_{\mathit{in}}$ & (\ref{eq:cin-e5m2-mul-rne}) & (\ref{eq:cin-e5m2-mul-rna}) & $0$ & (\ref{eq:cin-e5m2-mul-ru}) & (\ref{eq:cin-e5m2-mul-rd}) & $0$ & $0$\\
			$x^2$ 	   & $X \ll 1+\mathtt{0xc4} + c_{\mathit{in}}$ & $0$ & (\ref{eq:cin-e5m2-sq-rna}) & $0$ & (\ref{eq:cin-e5m2-sq-ru}) & $0$ & $0$ & $0$ \\
			$x/y$ 	   & $X-Y+\mathtt{0x3b} + c_{\mathit{in}}$ & (\ref{eq:cin-e5m2-div-rne-rna-rnz}) & (\ref{eq:cin-e5m2-div-rne-rna-rnz}) & (\ref{eq:cin-e5m2-div-rne-rna-rnz}) & (\ref{eq:cin-e5m2-div-ru}) & (\ref{eq:cin-e5m2-div-rd}) & (\ref{eq:cin-e5m2-div-rz}) & $0$ \\
			$1/x$ 	   & $-X+\mathtt{0x87} + c_{\mathit{in}}$ & (\ref{eq:cin-e5m2-recip-rne-rna-rnz}) & (\ref{eq:cin-e5m2-recip-rne-rna-rnz}) & (\ref{eq:cin-e5m2-recip-rne-rna-rnz})  & (\ref{eq:cin-e5m2-recip-ru}) & (\ref{eq:cin-e5m2-recip-rd}) &  (\ref{eq:cin-e5m2-recip-rz}) & $1$\tnote{b} \\
			$\sqrt{x}$ & $X\gg 1 + \mathtt{0x1e} + c_{\mathit{in}}$ & $0$ & $0$ & $0$ & (\ref{eq:cin-e5m2-sqrt-ru}) & --- & --- & $0$\\
			$1/\sqrt{x}$ & $-X \gg 1 + \mathtt{0x5a} + c_{\mathit{in}}$ & $0$ & $0$ & $0$ & (\ref{eq:cin-e5m2-sqrt-ru}) & --- & --- & $0$ \\
		\end{tabular}
		\smallskip
		\scriptsize
		\begin{tablenotes}
			\item[a] Where --- indicates that the rounding mode cannot be obtained using the integer expressions
			\item[b] Instead of having an unconditional $1$ on the carry in, the constant can be increased 
		\end{tablenotes}
		\end{threeparttable}
	\end{table*}
	\begin{table*}
		\centering
		\begin{threeparttable}
			\caption{Summary of Suggested Expressions for E4M3.}
			\label{tab:e4m3-expressions}
			\begin{tabular}{c|c|*{7}{c}}
				& & \multicolumn{7}{c}{Boolean expressions for carry in, $c_{\mathit{in}}$\tnote{a}} \\
				Operation   & Integer arithmetic expression & RN$_e$ & RN$_a$ & RN$_z$ & RU & RD & RZ & Faithful \\
				\hline
				$x\times y$ & $X + Y + \mathtt{0xc8} + c_{\mathit{in}}$ & (\ref{eq:cin-e4m3-mul-rne}) & (\ref{eq:cin-e4m3-mul-rna}) & (\ref{eq:cin-e4m3-mul-rnz}) & --- & --- & (\ref{eq:cin-e4m3-mul-rz}) & (\ref{eq:cin-e4m3-mul-faith}) \\
				$x^2$ 	   & $X \ll 1+\mathtt{0xc8}+ c_{\mathit{in}}$ & (\ref{eq:cin-e4m3-sq-rne-rnz}) & (\ref{eq:cin-e4m3-sq-rna}) & (\ref{eq:cin-e4m3-sq-rne-rnz}) & --- &  (\ref{eq:cin-e4m3-sq-rn-rz}) &  (\ref{eq:cin-e4m3-sq-rn-rz}) & (\ref{eq:cin-e4m3-sq-faith}) \\
				$x/y$ 	   & $X-Y+\mathtt{0x37} + c_{\mathit{in}}$ & (\ref{eq:cin-e4m3-div-rne-rna-rnz}) & (\ref{eq:cin-e4m3-div-rne-rna-rnz}) & (\ref{eq:cin-e4m3-div-rne-rna-rnz}) & --- & --- & --- & (\ref{eq:cin-e4m3-div-faith}) \\
				$1/x$ 	   & $-X+\mathtt{0x6f} + c_{\mathit{in}}$ & (\ref{eq:cin-e4m3-recip-rne-rna-rnz}) &  (\ref{eq:cin-e4m3-recip-rne-rna-rnz}) &  (\ref{eq:cin-e4m3-recip-rne-rna-rnz})  & --- & --- & --- & (\ref{eq:cin-e4m3-recip-faith}) \\
				$\sqrt{x}$ & $X \gg 1 +\mathtt{0x1b}+ c_{\mathit{in}}$ & (\ref{eq:cin-e4m3-sqrt-rne-rna-rnz}) & (\ref{eq:cin-e4m3-sqrt-rne-rna-rnz}) & (\ref{eq:cin-e4m3-sqrt-rne-rna-rnz}) & --- & (\ref{eq:cin-e4m3-sqrt-rn-rz}) & (\ref{eq:cin-e4m3-sqrt-rn-rz}) & $0$\\
				$1/\sqrt{x}$ & $-X \gg 1 + \mathtt{0x53}+ c_{\mathit{in}}$ & (\ref{eq:cin-e4m3-recip-sqrt-rne-rna-rnz}) & (\ref{eq:cin-e4m3-recip-sqrt-rne-rna-rnz}) & (\ref{eq:cin-e4m3-recip-sqrt-rne-rna-rnz}) & --- & (\ref{eq:cin-e4m3-recip-sqrt-rn-rz}) & (\ref{eq:cin-e4m3-recip-sqrt-rn-rz}) & 1\tnote{b} \\
			\end{tabular}
			\smallskip
			\scriptsize
			\begin{tablenotes}
			\item[a] Where --- indicates that the rounding mode cannot be obtained using the integer expressions
				\item[b] Instead of having an unconditional $1$ on the carry in, the constant can be increased 
			\end{tablenotes}
		\end{threeparttable}
	\end{table*}

	\section{Hardware Implementation of Multiplication}\label{sec:hw-implementation}
	To demonstrate the potential savings of these schemes, a minimal example for floating-point multiplication is implemented and synthesized for both ASIC and FPGA.
	Two different rounding are considered, the first one being RN$_e$ as it most commonly used, and the second one being RZ since it can be considered one of the cheapest rounding modes.
	
	Note, as multiplication is the simplest circuits of the considered operations to implement for floating-point, next to square, greater improvements can likely be obtained for the other operations. 
	
	As a reference, a multiplier based on the one presented in\cite{Muller2009Handbook} is used, but with the difference that the calculation of ${E_x+E_y-b}$ and ${E_x+E_y-b + 1}$ is not done in parallel in order to save area. Also, the reference multiplier only supports normalized operands and no handling of NaN or inf to have a fair comparison.
	
	There is also a combined reference multiplier, which can be seen as a E5M3 multiplier, but with configurable rounding and bias. However, as the results are worse for this than by simply multiplex the outputs of the two dedicated variants, there is also a version which does exactly that. Additional efforts will be spent on trying to optimized the combined multiplier for the final version.
	
	For the final version, energy figures based on switching-activity simulations will be provided for the standard cell implementations.
	
	\subsection{ASIC Results}
	The synthesis was made using Synopsys Design Compiler with a $28$~nm FD-SOI standard cell library at $1.0$~V power supply voltage. A  slow-slow process corner at an ambient temperature of $125~^\circ$C was considered to keep the results on the pessimistic side.
	Registers are added on the input and output to obtain a more correct critical path, but the registers are not included in the results.
	
\begin{figure}
		\centering
		\includegraphics[]{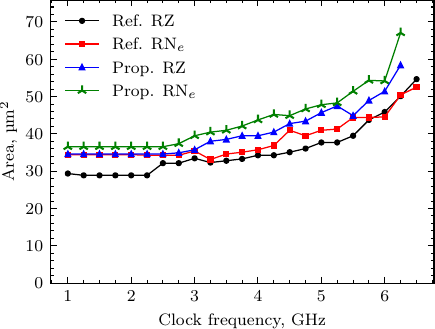}
		\caption{Area consumption for E5M2 multipliers.}
		\label{fig:e5m2-asic-area}
	\end{figure}

\begin{figure}
		\centering
		\includegraphics[]{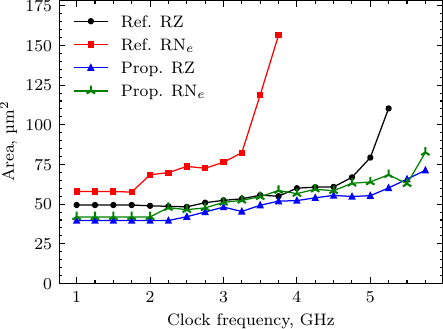}
		\caption{Area consumption for E4M3 multipliers.}
		\label{fig:e4m3-asic-area}
	\end{figure}

\begin{figure}
	\centering
	\includegraphics[]{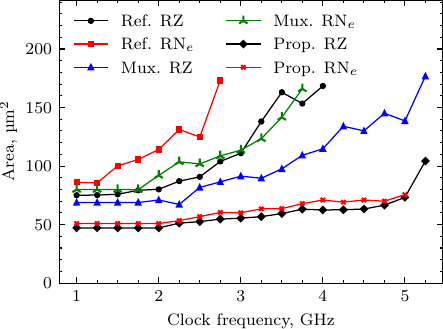}
	\caption{Area consumption for combined E4M3/E5M2 multipliers.}
	\label{fig:combined-asic-area}
\end{figure}

	Figures~\ref{fig:e5m2-asic-area},~\ref{fig:e4m3-asic-area}, and~\ref{fig:combined-asic-area} show the area consumption of the designs with respect to clock frequency. For E5M2, the proposed designs are slightly larger compared to the reference\footnote{However, using other commercial synthesis tools, improvements are present for the proposed over the reference.}. 
	Clear improvements are seen for E4M3, both in terms of area and maximum speed.
	Notably, the reference circuit using RN$_e$ reached $3.75$~GHz with an area of $160$~\textmu m$^2$, while the proposed designed reached $5.75$~GHz using only and area of $85$~\textmu m$^2$.
	
	Finally, for the combined multiplier, significant savings are again present for the proposed approach. Somewhat surprisingly, it is more efficient to simply multiplex the outputs of two dedicated multipliers rather than using an approach with a E5M3 multiplier. As mentioned earlier, these results will be further investigated to find a more efficient implementation for the final version.
	
	\subsection{FPGA Results}
	The design was synthesized for an AMD Artix-$7$, \texttt{xc7a12tcpg238-3}. Registers are again to the input and out, hence $24$ flip-flops were used in all designs ($25$ for the combined as there is an additional register for the control signal).
	The synthesis results can be found in Table~\ref{tab:fpga-results}.
	
		\begin{table}
		\caption{FPGA Synthesis Results for Multipliers.}
		\label{tab:fpga-results}
		\centering
		\begin{tabular}{  c c | c c }
			Format & Implementation & Delay, ns & LUTs \\
			\hline
			\multirow{4}{*}{E5M2} & Prop. RN$_e$   & $1.775$ & $8$  \\ 
			& Prop. RZ	   & $1.607$ & $8$  \\ 
			& Ref. RN$_e$    & $2.581$ & $10$ \\ 
			& Ref. RZ	       & $2.342$ & $10$ \\ 
			\hline
			\multirow{4}{*}{E4M3} &	Prop. RN$_e$   & $2.575$ & $8$  \\ 
			&	Prop.  RZ	   & $2.297$ & $8$  \\ 
			&	Ref. RN$_e$    & $4.318$ & $18$ \\ 
			&	Ref. RZ 	   & $3.458$ & $17$ \\ 
			\hline
			\multirow{4}{*}{Combined} &	Prop. RN$_e$   & $3.155$ & $9$  \\ 
			&	Prop.  RZ	   & $2.308$  &  $9$ \\ 
			&	Ref. RN$_e$    & $5.024$ & $44$ \\ 
			&	Ref. RZ 	   & $4.143$ & $36$ \\ 
			& Mux. RN$_e$ & $4.300$ & $30$ \\ 
			& Mux. RZ & $4.153$ & $27$ \\ 
		\end{tabular}
	\end{table}
	
	For the E5M2 case, fewer LUTs are used and the critical path is reduced.
	The proposed design of the E4M3 multiplier is again significantly better than its reference counterpart.
	For both rounding modes, the LUT count is more than halved. The delay is decreased by about $40\%$ for RN$_e$ and about $33\%$ for RZ.
	
	Hence, the rather long expressions in (\ref{eq:cin-e4m3-mul-rne}) and (\ref{eq:cin-e4m3-mul-rz}) can be mapped to the same 6-input LUT required for the addition of the least significant bits. As many FPGAs support ternary addition of three general numbers using a single LUT per bit \cite{simkins2007structures,kumm2018scm}, it is not surprising that they also support addition of two general numbers and a constant. Hence, one may expect that all other computations also will fit within eight 6-input LUTs.
	
	For the combined multiplier, again, significant savings are obtained using the proposed approach. The carry-in generation now cannot fit within a 6-input LUT as the mode select signal is also required. On the other hand, the complexity of reference multiplier increases significantly. For FPGA, it is clearly more efficient to simply multiplex the outputs of the two reference multipliers.
	
	
	
	\section{Conclusions} \label{sec:conclusions}
	In this work, eight-bit floating-point operations using simple integer operations is discussed.
	It is shown that correct rounding can be obtained in most cases for the considered operations when using a two-bit mantissa format, and faithful rounding can always be obtained.
	Faithful rounding can sometimes be obtained directly when using a three-bit mantissa format. However, by using a conditional carry in, correct rounding, and several other rounding modes, can be obtained for all considered operations and formats.
	Boolean expressions for the carry in are presented for all cases where it can be used. Results for hardware implementation of multipliers are presented, synthesized to both standard cells and FPGA. For E5M2 on FPGA and for E4M3 on both platforms, significant complexity savings are obtained. In addition, combined multipliers supporting both formats are implemented and, again, significant savings are obtained using the proposed approach.

	\bibliographystyle{IEEEtran}
	\bibliography{../IEEEabrv,../abbreviations/DAabrv,ref}
	
\end{document}